\documentclass[twocolumn]{aastex631}
\usepackage{booktabs}
\usepackage{gensymb}
\usepackage{upgreek}
\usepackage{multirow}
\usepackage{bigdelim}

\newcommand{\kmss}{km~s$^{-1}\ $}

\newcommand{\Lyas}{Ly$\alpha$ }
\newcommand{\HIs}{\ion{H}{1} }
\newcommand{\SiIIs}{\ion{Si}{2} }
\newcommand{\SiIIIs}{\ion{Si}{3} }
\newcommand{\CIIs}{\ion{C}{2} }
\newcommand{\SiIVs}{\ion{Si}{4} }
\newcommand{\CIVs}{\ion{C}{4} }
\newcommand{\NVs}{\ion{N}{5} }
\newcommand{\OVIs}{\ion{O}{6} }

\newcommand{\Ds}{\citet{Danforth2016}\ }

\newcommand{\M}{M$_{\odot}$}

\newcommand{\HI}{\ion{H}{1}}
\newcommand{\SiII}{\ion{Si}{2}}
\newcommand{\SiIII}{\ion{Si}{3}}
\newcommand{\CII}{\ion{C}{2}}
\newcommand{\SiIV}{\ion{Si}{4}}
\newcommand{\CIV}{\ion{C}{4}}
\newcommand{\NV}{\ion{N}{5}}
\newcommand{\OVI}{\ion{O}{6}}
\newcommand{\kms}{km~s$^{-1}$}

\newcommand{\D}{\citet{Danforth2016}}

\begin{document}

\title{Investigating Ionization in the Intergalactic Medium}

\shorttitle{Ionization in the IGM}
\shortauthors{Koplitz et al.}

\correspondingauthor{Brad Koplitz} 
\email{brad.koplitz@asu.edu}

\author[0000-0001-5530-2872]{Brad Koplitz}
\affil{School of Earth \& Space Exploration, 
Arizona State University, 781 Terrace Mall, Tempe, AZ 85287, USA}

\author[0000-0001-5155-8862]{Anjali Ramesh}
\affil{School of Earth \& Space Exploration, 
Arizona State University, 781 Terrace Mall, Tempe, AZ 85287, USA}

\author[0000-0002-2724-8298]{Sanchayeeta Borthakur}
\affil{School of Earth \& Space Exploration, 
Arizona State University, 781 Terrace Mall, Tempe, AZ 85287, USA}

\begin{abstract}

The Intergalactic Medium (IGM) contains $>$50\% of the baryonic mass of the Universe, yet the mechanisms responsible for keeping the IGM ionized has not been fully explained.
Hence, we investigate ion abundances from the largest blind QSO absorption catalog for clouds that show \CIV, \NV, and \OVIs simultaneously.
The wavelength range of present UV spectrographs, however, make it possible to probe \CIVs and \OVIs over a small range of redshift ($z\approx0.12-0.15$).
As a result, we only have five IGM absorbing clouds, yet these provide a powerful and representative tool to probe the IGM ionization state.
We found one cloud to be in collisional ionization equilibrium while three of five showed signs of being produced by non-equilibrium processes, specifically conductive interfaces and turbulent mixing layers.
None of the models we explore here were able to reproduce the ionization state of the remaining system.
Energetic processes, such as galactic feedback from star formation and AGN winds, would be excellent candidates that can cause such widespread ionization.

\end{abstract}


\section{INTRODUCTION}

Most of the baryonic matter in the universe is not contained in stars and galaxies, but is between galaxies in a dilute, multi-phase, ionized gas called the Intergalactic Medium (IGM; \citealt{Meiksin2009,McQuinn2016}).
This reservoir is thought to regulate the growth of galaxies by facilitating accretion (e.g., \citealt{Keres2005,Dekel2006,Dekel2009,Hafen2022,Decataldo2023}) and harboring a large fraction of what gets ejected through outflows (e.g., \citealt{Martin1999,Martin2010,Steidel2010,Peeples2014,Oppenheimer2016}).
Many studies of this diffuse gas have been done at intermediate redshifts ($z \approx 2 - 5$) to allow for the simultaneous detection of multiple Lyman transitions of hydrogen as well as metals such as \CIVs or \OVIs (e.g, \citealt{Bergeron1994,Jannuzi1998,Lopez1999,Richter2004,Simcoe2004,Adelberger2005,Chen2005,Danforth2008,Turner2014,Morrison2021,Borthakur2022}).

Metals, in particular, are an import tracer of this diffuse gas as 70\% of \Lyas forest absorbers are found to have accompanying metal lines \citep{Simcoe2004} and can be present even when the Lyman series is weak \citep{Danforth2016}.
The IGM at $z \approx 2 - 3$ is also known to be enriched with carbon and oxygen (e.g., \citealt{Dave1998,Aracil2004,Pieri2006}) and is thought to have retained it and other metals to the present day (e.g., \citealt{Richter2004,Aguirre2008,Danforth2008,Tripp2008,Muzahid2012,Danforth2016}).
Additionally, metal absorption lines are often unsaturated, allowing for more components within a single cloud to be detected (e.g., \citealt{Chen2009,Danforth2016,Pachat2017,Sankar2020,Ahoranta2021}).
This makes metals a key window into the ionization processes that govern the IGM.

Furthermore, analyzing metals has revealed the multiphase nature of the IGM (e.g., \citealt{Heckman2002,Savage2005,Narayanan2009,Shull2012,Ahoranta2021,Haislmaier2021}), with a cool $T \approx 10^{4.5}$~K phase and a warm-hot phase, known as the warm ionized IGM or WHIM, at $T \approx 10^5 - 10^6$~K.
The presence of the WHIM could indicate that collisions are likely a dominate ionization process in the IGM since many of these processes produces radiatively cooling gas at the intermediate temperatures \OVIs is found at (e.g., \citealt{Begelman1990,Heckman2002,Gnat2010,Kwak2015,Ji2019}), although multiple studies have assumed photoionization equilibrium (PIE) or collisional ionization equilibrium (CIE) to estimate IGM masses and densities (e.g., \citealt{Lehner2007,Sobacchi2013}).

In the largest, most complete IGM survey to date, \Ds identified 5138 extragalactic absorption features along 82 QSO sight lines taken with the Cosmic Origins Spectrograph (COS) aboard the Hubble Space Telescope (HST; \citealt{Osterman2011,Green2012}) with signal-to-noise ratios above 15.
They grouped features found at similar redshifts into 2611 absorbing systems.
16\% of these had at least one metal line, with \OVIs being most frequently detected, similar to what other studies have found in the local universe (e.g., \citealt{Danforth2008,Tilton2012}).
The number of absorbers per unit $z$ ($\frac{d\mathcal{N}}{dz}$) of \OVIs and \HIs were found to increase with $z$; however, this was not seen in \NV, \ion{C}{3}, or \SiIIIs which does not seem to evolve \citep{Danforth2016}.

Despite the progress that has been made, it is still unclear whether this diffuse gas is in ionization equilibrium $-$ photoionization or collisional ionization $-$ or if non-equilibrium processes are needed to explain observations.
To that end, we have analyzed the absorption features from \Ds which allow us to constrain the physical processes driving ionization in the IGM in the context of equilibrium and non-equilibrium interactions.
The rest of this paper is outlined as follows:
Section~\ref{sect:sample} details how our sample was selected as well as the measurements used in our analysis.
Section~\ref{sect:results} presents the analysis and the results from it.
Finally, we summarize in Section~\ref{sect:summary} and discuss future directions.

\section{Sample \& Measurements} \label{sect:sample}

\subsection{Sample Selection}

The presence of high ionization transitions can be used to constrain the physical nature of the IGM. 
In particular, warm-hot metals such as \CIVs ($\lambda\lambda1548,1550$~\AA; 64.5~eV), \NVs ($\lambda\lambda1238,1242$~\AA; 97.9~eV), and \OVIs ($\lambda\lambda1032,1038$~\AA; 138.1~eV), trace the energies required to ionize the IGM and provide a unique window into some of the commonly observed non-equilibrium processes that are most likely responsible for the ionization state of the IGM.
To that end, we consider all absorbing clouds (i.e., sets of absorption features which are aligned in velocity space) from \D's Mikulski Archive for Space Telescopes (MAST) catalog\footnote{\url{https://archive.stsci.edu/prepds/igm/}} and examine those with 3$\sigma$ detections of these three ions.
While it is possible that absorption features are kinematically aligned by chance and not actually associated with one another, the likelihood is very small when multiple ions are found at similar velocities.
As a result, we assume all kinematically aligned ions originate from a single absorbing cloud.

The selection criteria yield five clouds along four sight lines that showed absorption in all three transitions.
Each of these clouds were best fit by a single Voigt profile or component.
The sight line PG~1216+069 contained two clouds with a velocity separation of $\sim$75~\kms.
Although the \Ds identified them as independent, these could be associated with a larger structure.

Even though only five absorbing clouds from the \Ds catalog are included in our complete sample, these are representative of the IGM.
The small number can be attributed to the narrow $z$ range within which \CIVs and \OVIs can be simultaneously observed using COS.
At minimum, $z$ needs to be $\gtrsim$0.094 to have at least one \OVIs feature within the G130M grating, and $z \gtrsim 0.100$ to observe the stronger $\lambda1032$~\AA\ line.
Meanwhile, any cloud with $z \gtrsim 0.153$ will have both \CIVs features shifted out of the G160M grating.
Only 341 clouds were within the allowable range, limiting the number that could have been included.
Most of these only contained \Lyas with no associated metal features (246 of 341).
\OVI, the most frequently detected metal, appeared in 44 individual clouds ($\mathcal{N}$), implying $\frac{d\mathcal{N}}{dz}(\mathrm{O\ VI}) \approx 9$, which is comparable to what \citet{Danforth2008} found for \OVIs ($15^{+3}_{-2}$).
The fact that the 82 QSOs from \Ds are randomly distributed across the sky indicates that these absorbers are likely more prevalent than the small sample size would imply.

\subsection{Voigt Profile Measurements} \label{sect:voigt}

We fit Voigt profiles to absorption features to determine the column density ($N$), Doppler width ($b$), and relative velocity or velocity centroid ($v_\mathrm{obs}$) of the absorbing gas and check for consistency with the analysis from \D.
In addition, \Ds fit each absorption profile individually whereas we fit doublets, such as \CIVs and \OVI, simultaneously.
Before fitting, we normalized the continuum within $\pm$600~\kmss of the cloud's $z$, which we refer to as $z_\mathrm{sys}$, and center the features such that $v_\mathrm{obs}$ is always near 0~\kms.
The absorption features were fit using a reduced $\chi^2$ algorithm \citep{Sembach1992}.
Once complete, these Voigt profile parameters can be used to constrain the ionization processes happening in the IGM.
We show the spectrum and associated fits of this analysis for the 4 sightlines in Appendix.

We present the absorber properties used in our analysis in Table~\ref{tab:results}.
The $v_\mathrm{obs}$ returned by our fits are not reported since each was centered on $z_\mathrm{sys}$ before fitting.
If the best-fit Doppler width ($b$) of an absorber is narrower than 5~\kms, we fix $b$ to 5~\kmss and report the resulting column density ($N$).
This is motivated by the COS line spread function since it is unable to discern between $b$ values $\leq5$~\kmss at a signal to noise ratio between 10 and 20.
This is a limiation of the data.
It is possible for these clouds to be associated with extremely low Doppler width and much higher column density.
Most of our measurements were consistent with those from \Ds and so we adopt our values.
We also fit the \SiIVs ($\lambda\lambda1393,1402$~\AA; 45.1~eV) features of our clouds as they provide an additional probe of the ionization mechanisms; however, we do not include the ion in our selection criteria as it may not completely trace the WHIM as indicated by its low ionization potential \citep{Cen&Ostriker1999,Dave2001,Cen&Ostriker2006,McQuinn2016}.
The lower ions \SiIIs ($\lambda\lambda1260,1193,1190$~\AA; 16.4~eV), \CIIs ($\lambda1334$~\AA; 24.4~eV), and \SiIIIs ($\lambda1260$~\AA; 33.5~eV) were also fit when detected.
If no absorption is present, we measure twice the error of the rest-frame equivalent width within a 100~\kmss window that does not contain intervening absorption.
This is then converted to an upper limit on $N$, assuming we are in the linear portion of the curve of growth.
Neither absorbing cloud along the sightline PG~1216+069 contained \SiIIs or \CII, so we report the same upper limits for both clouds.

Any Voigt profile analysis is limited by the resolution of the spectrograph used.
The COS instrument currently has the highest spectral resolution at rest-frame FUV wavelengths that could observe the QSOs in the \Ds sample.
However, it is important to note that it is a medium resolution instrument ($R$~$\approx$~20,000; FWHM~$\approx$~15~\kms).
Thus, it is possible that multiple narrow clouds at similar velocities could appear as a single, wider component.
We assume that the measurements presented here are dominated by the largest absorbing cloud with the understanding that higher resolution observations may reveal a more complicated picture.

\begin{deluxetable}{ccccc}
\centering
\tablecolumns{5}
\tablewidth{0pt}
\setlength{\tabcolsep}{3.5pt}
\label{tab:results}
\tablecaption{Summary of Measurements}
\tablehead{
\colhead{QSO} & \colhead{$z_\mathrm{sys}$} & \colhead{Ion} & \colhead{log$(N/\mathrm{cm}^{-2})$} & \colhead{$b \left[\mathrm{km\ s}^{-1}\right]$} \\
\colhead{(1)} & \colhead{(2)} & \colhead{(3)} & \colhead{(4)} & \colhead{(5)}}
\startdata
 \multirow{8}{*}{(1) PG 1116+215} & \multirow{8}{*}{0.13853} & \HI & $15.30 \pm 0.03$ & $28.9 \pm 1.0$ \\
 &  & \SiII & $12.93 \pm 0.04$ & $\leq 5^{\star}$ \\
 &  & \SiIII & $13.73 \pm 1.31$ & $5.8 \pm 1.9$ \\
 &  & \CII & $13.98 \pm 0.13$ & $8.6 \pm 1.3$ \\
 &  & \SiIV & $12.83 \pm 0.20$ & $5.9 \pm 4.2$ \\
 &  & \CIV & $13.40 \pm 0.09$ & $\leq 5^{\star}$ \\
 &  & \NV & $12.78 \pm 0.06$ & $15.0 \pm 1.2$ \\
 &  & \OVI & $13.82 \pm 0.02$ & $30.8 \pm 1.1$ \\
 \hline
\multirow{8}{*}{(2) PG 1216+069} & \multirow{8}{*}{0.12360} & \HI & $14.63 \pm 0.05$ & $22.8 \pm 1.1$ \\
 &  & \SiII & $<$ 11.88 & --- \\
 &  & \SiIII & $12.53 \pm 0.21$ & $7.3 \pm 2.9$ \\
 &  & \CII & $<$ 12.93 & --- \\
 &  & \SiIV & $12.71 \pm 0.11$ & $19.5 \pm 2.6$ \\
 &  & \CIV & $14.24 \pm 0.24$ & $10.4 \pm 1.3$ \\
 &  & \NV & $13.42 \pm 0.09$ & $\leq 5^{\star}$ \\
 &  & \OVI & $14.19 \pm 0.05$ & $20.2 \pm 1.4$ \\
 \hline
\multirow{8}{*}{(3) PG 1216+069} & \multirow{8}{*}{0.12389} & \HI & $14.86 \pm 0.06$ & $24.96 \pm 1.1$ \\
 &  & \SiII & $<$ 11.88 & --- \\
 &  & \SiIII & $12.51 \pm 0.10$ & $\leq 5^{\star}$ \\
 &  & \CII & $<$ 12.93 & --- \\
 &  & \SiIV & $12.84 \pm 0.12$ & $\leq 5^{\star}$ \\
 &  & \CIV & $13.99 \pm 0.15$ & $11.0 \pm 1.3$ \\
 &  & \NV & $13.04 \pm 0.12$ & $10.4 \pm 13.6$ \\
 &  & \OVI & $14.19 \pm 0.06$ & $32.0 \pm 1.4$ \\
 \hline
\multirow{8}{*}{(4) PG 1424+240} & \multirow{8}{*}{0.14713} & \HI & $15.56 \pm 0.12$ & $27.5 \pm 1.1$ \\
 &  & \SiII & $<$ 11.78 & --- \\
 &  & \SiIII & $14.31 \pm 1.21$ & $5.4 \pm 1.7$ \\
 &  & \CII & $<$ 12.82 & --- \\
 &  & \SiIV & $12.99 \pm 0.16$ & $7.8 \pm 2.7$ \\
 &  & \CIV & $14.24 \pm 0.16$ & $14.4 \pm 1.3$ \\
 &  & \NV & $13.16 \pm 0.16$ & $33.6 \pm 1.9$ \\
 &  & \OVI & $14.07 \pm 0.43$ & $\leq 5^{\star}$ \\
\hline
\multirow{8}{*}{(5) PKS 0637-752} & \multirow{8}{*}{0.12288} & \HI & $15.36 \pm 0.40$ & $30.5 \pm 1.2$ \\
 &  & \SiII & $<$ 11.74 & --- \\
 &  & \SiIII & $14.05 \pm 1.18$ & $8.2 \pm 1.7$ \\
 &  & \CII & $<$ 12.86 & --- \\
 &  & \SiIV$^{0}$ & --- & --- \\
 &  & \ldelim\{{3}{*}\CIV & $13.65 \pm 0.04$ & $38.8 \pm 1.2$ \\
 &  & \ \ \NV & $14.05 \pm 0.04$ & $71.9 \pm 1.1$ \\
 &  & \ \ \OVI & $14.08 \pm 0.03$ & $44.0 \pm 1.1$ \\ 
\enddata
\tablenotetext{}{Column~(1) shows the quasar that the system is towards. Column~(2) is the redshift of the system. Column~(3) indicates the ion being fit. Columns~(4) and (5) are the column density and Doppler width of the ion in units of cm$^{-2}$ and \kms, respectively. $^{\star}$ Doppler width was fixed at 5 \kmss due to the absorption line being narrow. See Section \ref{sect:voigt} for details. It is worth noting that these features are consistent with extremely narrow ($b << 5$~\kms) features implying large column densities. Measurements that are consistent with CIE models \citep{Gnat2007} are highlighted with a bracket. $^0$ Both transitions fell within a gap in the data, so a fit could not be completed.}
\end{deluxetable}

\subsection{Comparison to Literature Measurements}

Many of the clouds we analyze here have been identified by other studies and fit many of the same transitions as us.
We compare our measurements to those in the literature below.

In the system along PG~1116+215, \citet{Sembach2004}, \citet{Tripp2008}, \citet{Tilton2012}, and \citet{Muzahid2018} found similar values of log$N$ and $b$ to what we measure for most metals.
However, \citet{Tilton2012} found larger log$N$ values for \HIs and \NVs than our fits ($\sim$0.4 and $\sim$1 dex more, respectively).
\citet{Muzahid2018} found log$N$ of \HIs to be larger than we report while finding less \CIV. 
The fit from \citet{Sembach2004} gave a larger log$N$ and smaller $b$ than us.
Meanwhile, \citet{Tripp2008} fit the \HIs absorption as two components, both of which have narrower $b$ and larger log$N$ values than we find.

As discussed in Section~\ref{sect:sample}, the small velocity separation of the two absorbing clouds towards PG~1216+069 could cause them to be treated as one system which is what \citet{Tripp2008}, \citet{Chen2009}, and \citet{Tilton2012} have done.
\citet{Tripp2008} found values for \OVIs which are consistent with what we measure.
Their \HIs measurements for the system at $z = 0.12360$ are consistent with what we find, though they find a larger log$N$ and smaller $b$ than we do for system 3 at $z = 0.12389$.
The \OVIs features measured by \citet{Chen2009} have similar, but inconsistent, log$N$ as we find while their $b$ value of the system at $z = 0.12360$ is narrower.
For \SiIII, \citet{Chen2009} set $b = 2.4$ \kmss and found the corresponding best fit log$N$, which is consistent with what we find.
The \HIs features of both clouds were reported as lower limits which is consistent with the system at $z = 0.12360$ but not the system at $z = 0.12389$; however, Ly$\beta$ was not included in their fits which may have impacted this.
\citet{Tilton2012} measured similar log$N$ and $b$ values for \OVIs in both systems.

For the PG~1424+240 system, \citet{Muzahid2018} measured log$N$ values for \CIVs and \OVIs that are consistent with what we find.
However, they measured larger values of log$N$ for \HI, \CII, and \SiIIs than we do.
Part of the difference in \HIs can likely be attributed to how the values are reported.
They report a total log$N$ of \HIs from a fit with four components while we fit the feature with two components and only report the value of the single cloud which is kinematically aligned with the associated metal lines. 
The discrepancy for the \CIIs and \SiIIs features can be attributed to how the lines were identified. 
\citet{Danforth2016} labeled these as weak \HIs absorbing clouds, and so we report an upper limit for these metals where \citet{Muzahid2018} fit Voigt profiles.

In the PKS~0637$-$752 system, \citet{Johnson2017} fit two components to the \HI, \CIV, and \OVIs features and one component to \SiIIIs whereas we fit a single component to all of the metals features and two components for \HI.
This difference caused our $b$ values of \CIVs and \OVIs to be larger than what they find, though we measured similar total log$N$ values for all metals.
Our log$N$ value for \HIs is consistent with what they find, though we find a larger $b$ value.

\section{Results \& Discussion} \label{sect:results}

\subsection{Collisional Ionization Equilibrium Models} \label{sect:cie}

\begin{figure}
    \centering
    \includegraphics[width=\linewidth]{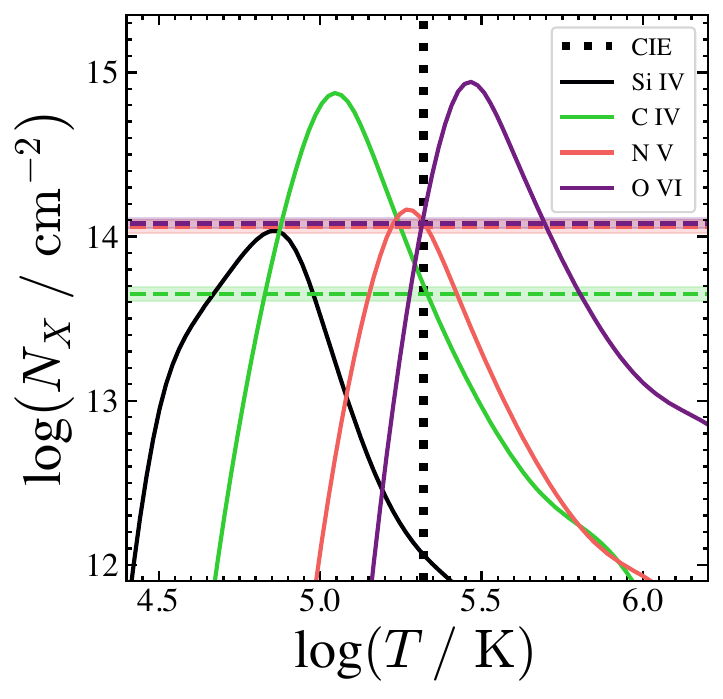}
    \caption{Comparing our measurements of the PKS~0637-752 cloud to CIE models \citep{Gnat2007}. Measurements are shown as dashed lines with uncertainties as shaded regions while CIE models are shown as solid lines. The color indicates the ion being plotted. Black for \SiIV, green for \CIV, red-orange for \NV, and purple for \OVI. Both \SiIVs lines fell within a gap in the data and so a measurement could not be made. The black dotted line shows where the cloud is in CIE at $T \approx 10^{5.3}$~K.}
    \label{fig:cie}
\end{figure}

To determine whether the absorbing clouds we have identified are in collisional ionization equilibrium (CIE), we compare the measurements to the solar metallicity \citep{Asplund2009} models from \citet{Gnat2007}.
PKS~0637$-$752 is the only sight line that is consistent with CIE at a temperature near 10$^{5.3}$~K, which we show in Figure~\ref{fig:cie}.
The remaining four clouds had less \NVs than predicted.
\citet{Gatuzz2023} looked at the cloud along the PG~1116+215 sight line and concluded that it was not in CIE, consistent with what we find here.

\subsection{Photoionization Equilibrium Models} \label{sect:pie}

The photoionization equilibrium (PIE) code \texttt{CLOUDY} (v.17;~\citealt{Ferland2017}) allows us to explore whether our measurements can be explained by an incident radiation field alone.
Each model was exposed to a \citet{HM2012} extragalactic UV background and was iterated until the Ly$\alpha$ column density was reached.
The metallicities of the features in our sample are assumed to be solar \citep{Asplund2009}.
To determine the total H density (log~$n_\mathrm{H}$) of the clouds, we varied log~($n_\mathrm{H}$/cm$^{-3}$) in steps of 0.1~dex from $-7$ to $-2$.

\begin{figure}
    \centering
    \includegraphics[width=\linewidth]{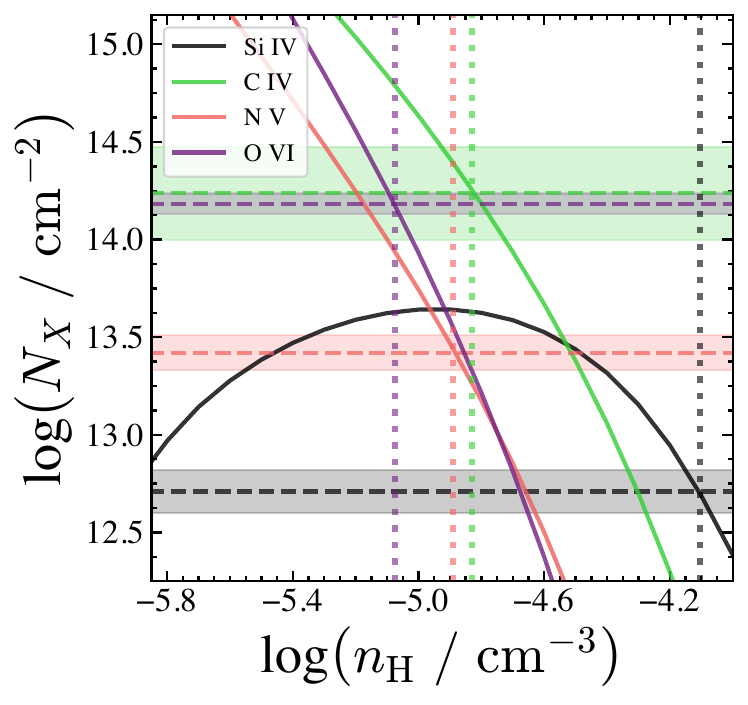}
    \caption{Comparing our measurements of the PG~1216+069 absorbing cloud at $z = 0.12360$ (system 2) to PIE models \citep{Ferland2017}. Measurements are shown as horizontal dashed lines with uncertainties as shaded regions while PIE models are shown as solid lines. The log $n_\mathrm{H}$ values which best matches our measurements are highlighted with vertical dotted lines. The color indicates the ion being plotted. Black for \SiIV, green for \CIV, red-orange for \NV, and purple for \OVI.}
    \label{fig:pie}
\end{figure}

No absorbing cloud was found to be consistent with PIE models, regardless of the log~$n_\mathrm{H}$ used.
We show the measurements of the absorbing cloud towards PG~1216+069 at $z = 0.12360$ in Figure~\ref{fig:pie} as an example.
This shows that there is no density consistent with the measured values for all four ions from these models (log~($n_\mathrm{H}$/cm$^{-3}$) $\approx$ $-$4.1, $-$4.8, $-$4.9, and $-$5.1 for \SiIV, \CIV, \NV, and \OVIs respectively). 
So the absorbing cloud is not consistent with being in PIE.
Changing the metallicity of the models would not impact this conclusion as this would move all of the models up or down together concurrently and would not impact their ratios.
In four of five absorbing clouds, \CIVs and \OVIs were found in similar amounts, which \texttt{CLOUDY} was not able to match without needing significantly more \NVs than we measured.
These show that photoionization is not the dominant ionization process for most high ions, unlike previously believed (e.g., \citealt{Narayanan2009,Muzahid2011}).

\subsection{Non - Equilibrium Models} \label{sect:noneq}

\begin{figure*}
    \centering
    \includegraphics[width=\linewidth]{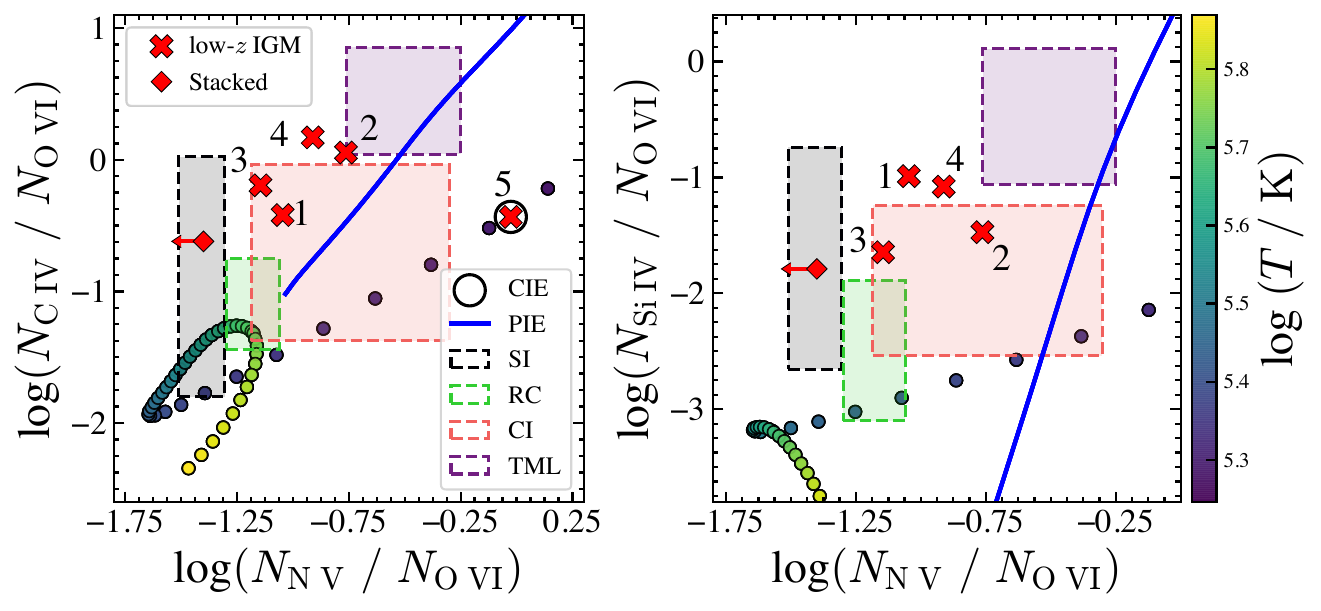}
    \caption{Comparing our measurements to various ionization models. We show our measurements for the 5 clouds as red stars while the stacked spectra is shown as a red diamond, which we discuss in Section~\ref{sect:stack}. The measurement uncertainties were always smaller than the size of the points and so are not shown. Each cloud is label with their corresponding number from Table~\ref{tab:results}. CIE models are shown as circles with the color corresponding to the model's temperature as shown in the color bar. The black solid ring highlights the cloud towards PKS~0637-720 which is consistent with being in CIE. PIE models are shown as a solid blue line. The dashed lines indices where the different non-equilibrium ionization models are expected to reside with the coloring indicating the model. Black for SI, green for RC, red-orange for CI, and purple for TML. These models are discussed in Section~\ref{sect:noneq}.}
    \label{fig:ionize_all}
\end{figure*}

Since only one cloud is consistent with an equilibrium model, we investigate our measurements for consistency with non-equilibrium models.
Figure~\ref{fig:ionize_all} shows the non-equilibrium models as well as the CIE and PIE models previously discussed.
Below, we detail each model while comparing the expected column density ratios for solar metallicity and relative abundances to our results.

Shock ionization (SI) can occur when gas clouds move through an ambient medium at velocities above the local sound speed.
In the IGM, this can be the result of galactic outflows giving clouds enough energy to escape the galaxy's dark matter gravitational potential.
When this happens, the temperature gets raised behind the cloud which causes higher ionization states to be populated, which \citet{Dopita1996} has modeled using clouds with varying shock velocities (150 and 500 \kms) and magnetic parameters (0 $\mu$G cm$^{-3/2}$ $\leq$ $B_0~n_0^{-3/2}$ $\leq$ 4 $\mu$G cm$^{-3/2}$).
These models are shown in Figure~\ref{fig:ionize_all} in black.
Most of our systems (four of five) have \CIV:\OVIs ratios consistent with the predicted values.
However, in each case we see significantly more \NVs than what is expected based on the models.

Radiative cooling (RC) via recombination can produce warm-hot ions as the gas temperature decreases from $>$10$^6$~K.
By cooling gas under a variety of conditions, \citet{Edgar1986} was able to predict the column density ratios one would see if RC is the dominate ionization method. 
The RC models for flow velocities of 100 \kmss are shown in green in Figure~\ref{fig:ionize_all}.
These ratios over predict the amount of \OVIs we see in all of the clouds.
In two clouds, we find nearly an order of magnitude more \CIVs than expected from the models.
As a result, we conclude that RC is not a prominent ionization mechanism in these absorbing clouds.

Conductive interfaces (CI) occur when media at different temperatures come into contact with each other; for example, when a cool cloud from a galaxy's interstellar medium gets ejected into the warm-hot IGM.
Collisions at the contact surface will transfer energy (and temperature) to the colder gas, producing the warm-hot metals we are interested in. 
\cite{Borkowski1990} modeled the contact surface between hot 10$^6$~K gas and cooler interstellar clouds, with the results being shown in Figure~\ref{fig:ionize_all} in red-orange.
These models vary the angle of the magnetic field orientation between 0$\degree$ and 85$\degree$ as well as interface ages between $10^5$ and $10^7$ yr. 
This age range suggests that these interfaces die out rather rapidly. 
Star formation driven galactic winds in starburst galaxies are known to produce high-ionization transitions such as \CIVs and \OVIs in the outer CGM and IGM \citep{Adelberger2005,Borthakur2013,Heckman2017,Mendez-Hernandez2022,Banerjee2023}, although the non-equilibrium nature of those systems are not fully explored.
Only system 3 (the cloud at $z = 0.12389$ towards PG~1216+069) is consistent with the predicted ratios in both panels of Figure~\ref{fig:ionize_all}.
Meanwhile, system 1 (the cloud towards PG~1116+215) and system 2 (the cloud at $z = 0.12360$ towards PG~1216+069) are only consistent with the left and right panels, respectively, though system 2 is near the boarder in the left panel.
The PKS~0637$-$752 cloud is in CIE and has more \NVs than the models predict, suggesting it may have once had a CI that has since dissipated as the cloud reached CIE.
Given that the ionization potential of \SiIVs allows it to be produced by the colder IGM gas phases than \CIVs or the hotter ions, we believe that the \CIVs measurements are a more robust tracer of the WHIM phase in the IGM.
As a result, we conclude that the ionization of systems 1, 2, and 3 can largely be explained by CIs resulting from energetic processes impacting the IGM.

When a turbulent hot gas comes into contact with a colder medium, Kelvin-Helmholtz instabilities can form, causing a mixing of different gases which are referred to as Turbulent Mixing Layers (TMLs).
These layers are at high enough temperatures to produce the highly ionized species we analyze here.
\citet{Slavin1993} expanded upon \citet{Begelman1990} to produce a model of TMLs over a range of temperatures and gas velocities.
These models, which we show in purple in Figure~\ref{fig:ionize_all}, correspond to entrainment velocities between 25 and 100 \kmss as well as temperatures between $10^{5.0}$ and $10^{5.5}$~K.
These models stand out amongst those we analyze because of the large \CIV:\OVIs and \SiIV:\OVIs ratios predicted.
The cloud along PG~1216+069 at $z = 0.12360$ (system 2) is near the boundary of TML in the left panel, suggesting this is a major contributor to its ionization state in addition to CIs.

The cloud towards PG~1424+240 (system 4) does not match the predicted values for any of the models we explore here.
Though it is near the boarder of CIs and TMLs in both panels of Figure~\ref{fig:ionize_all}.
This could indicate that these mechanisms are playing some role in the observed ionization state; however, there could other processes impacting these ionic ratios.
The uncertainty of the \OVIs log$N$ may also play a part in its positional inaccuracy on Figure~\ref{fig:ionize_all} as discussed in Section~\ref{sect:voigt}.

To summarize these results, most of the clouds we analyze are consistent with CIs when comparing \CIVs measurements.
Of which, one also match the predicted ratios for TMLs.
This is similar to what was found for Milky Way HVCs \citep{Fox2005}.
The models we investigate here are not able to reproduce the observed ratios of one system.
These results indicate that the IGM may be predominantly ionized through non-equilibrium processes.
More work is needed with a larger sample, however, to be able to draw more definitive conclusions.

These results could change if the relative abundances are largely different from solar values.
Though this is unlikely to be the case in the absorbing clouds we investigate here given the low redshifts they reside at and no study has found strong evidence for non-solar relative abundances in the IGM or outer CGM.
Additionally, \SiIVs can be produced by multiple gas phases, not just the warm-hot phase.
This means the placement in the right panel of Figure~\ref{fig:ionize_all} can be thought of as upper limits in the $y$-axis.
The multi-phase nature of \SiIVs does not impact the conclusions drawn here, given that the vast majority of the ionic abundances would need to be produced by cooler gas phases to change our interpretations.

\subsection{Possible Sources of Ionization}

With it being apparent from the above results that non-equilibrium processes are the dominate way in determining the ionization state of the IGM, it is important to look for the sources driving the non-equilibrium processes.
Galaxies with large outflows are capable of driving clouds out of the galaxy to the CGM and IGM \citep{Oppenheimer2012,Somerville&Dave2015}.
This is why using a galaxy's virial radius as boundary between the CGM and IGM is often insufficient or misleading, given that processes which occur at or near this border are likely to persist to further radii with little changing (e.g., \citealt{Nelson2019}).
To determine whether or not there are galaxies near the sight lines that could be responsible for the ionization processes we infer in Section~\ref{sect:cie}, we have performed a literature review of the absorbing clouds in our sample with the results summarized below and in Table~\ref{tab:sources}.
The Sloan Digital Sky Survey and other large spectroscopic surveys are shallow at these redshifts.
Thus, deeper individual surveys are needed to identify nearby galaxies.

The sight lines PG~1116+215, PG~1216+069, and PG~1424+240 were found to pass within 140~kpc \citep{Tripp2008,Muzahid2018,Scott2021}, 94~kpc \citep{Chen2009,Scott2021}, and 132~kpc \citep{Scott2021}, respectively, of galaxies at similar redshifts as these clouds.
The virial radii of these galaxies, as shown in Table~\ref{tab:sources}, suggest that the sight lines PG~1116+215 and PG~1216+06 probe the boundary between the outer CGM and IGM.
Given that CIs are shortly lived processes, these galaxies are likely responsible for sourcing the observed ionization in these clouds.
The sight line PG~1424+240 is the only cloud located outside the nearby galaxy's virial radius.
This suggests that the absorbing cloud escaped the gravitational potential of the galaxy and interacted with the ambient IGM to produce the observed ionization.

The PKS~0637$-$752 sight line stands out as it is only $\sim$16~kpc from a galaxy at the same redshift as the cloud in our sample, putting it into the inner CGM of this star-forming dwarf galaxy (M$_\star \approx 10^{7.9}$~\M; \citealt{Johnson2017}).
It is especially note worthy that this is the only cloud found to be in CIE at a temperature of $\sim$10$^{5.3}$~K.
This temperature is close to the virial temperature of this galaxy (T $\approx 10^{5.1}$~K), assuming a stellar mass to halo mass conversion of \citet{Kravtsov2013} and virial temperature estimation as described in \citet{Wang2008}.
In this case, we are most likely observing the virialized halo of this galaxy, although heated gas from feedback processes cannot be ruled out.

\begin{deluxetable}{cccccc}
\centering
\tablecolumns{6}
\tablewidth{0pt}
\setlength{\tabcolsep}{2pt}
\label{tab:sources}
\tablecaption{Summary of Possible Ionization Sources}
\tablehead{
\colhead{Sight line} & \colhead{$z_\mathrm{abs}$} & \colhead{$z_\mathrm{gal}$} & \colhead{R$_\mathrm{vir}$} & \colhead{$\rho$} & \colhead{Process(es)} \\
\colhead{(1)} & \colhead{(2)} & \colhead{(3)} & \colhead{(4)} & \colhead{(5)} & \colhead{(6)}}
\startdata
PG~1116+215 & 0.13853 & 0.138 & 172 & 140 & CI \\
PG~1216+069 & 0.12360,0.12389 & 0.124 & 125 & 94 & CI, TML \\
PG~1424+240 & 0.14713 & 0.15 & 81 & 132 & --- \\
PKS~0637$-$752 & 0.12288 & 0.1229 & 70 & 16 & CIE \\
\enddata
\tablenotetext{}{Column~(1) indicates the sight line. Columns~(2) and (3) are the redshift of the absorbing cloud and the galaxy, respectively. Column~(4) is the virial radius of the foreground galaxy in units of kpc. Column~(5) is the projected distance between the QSO and galaxy in units of kpc. Column~(6) shows the ionization processes consistent with the log$N$ ratios in Table~\ref{tab:results}. Acronyms refer to Conductive Interfaces (CI), Turbulent Mixing Layers (TML), and Collisional Ionization Equilibrium (CIE).}
\end{deluxetable}

\subsection{Stacked Spectra} \label{sect:stack}

\begin{figure*}
    \centering
    \includegraphics[width=\linewidth]{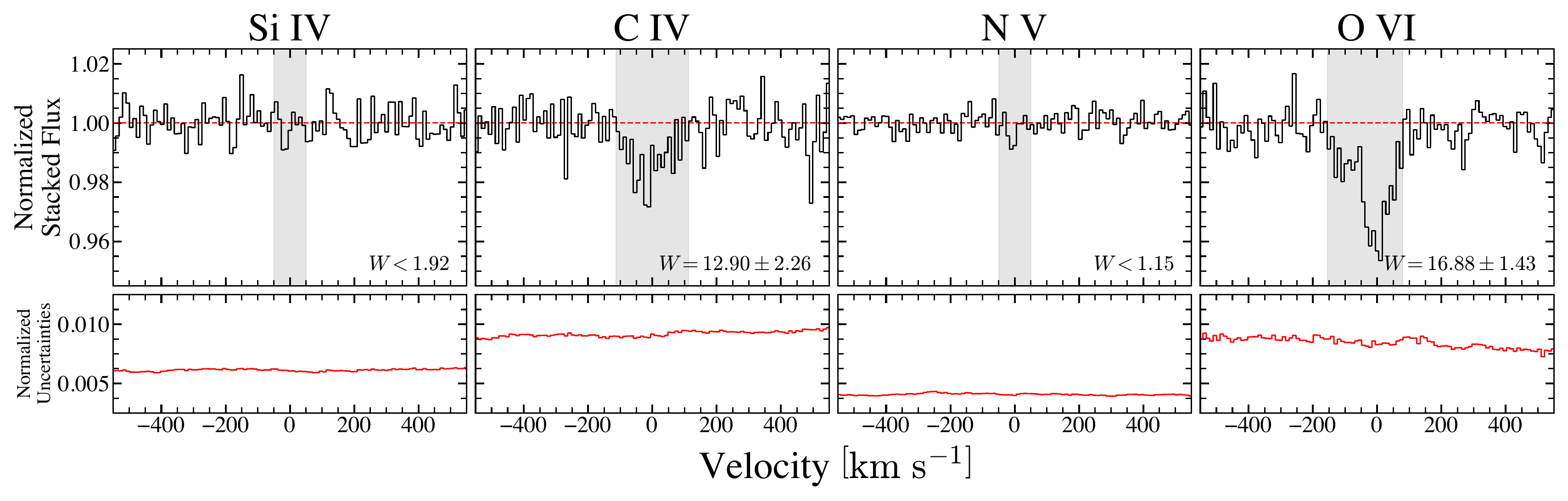}
    \caption{Stacked spectra for the expected location of \SiIVs (left) \CIVs (center left), \NVs (center right), and \OVIs (right) transitions for all IGM clouds in \D. The upper panels show the stacked flux while the bottom panels show the correspond to the associated stacked error. The grey region indicates the range used to calculate equivalent width, which we show in the bottom right of the upper panels in units of m\AA.}
    \label{fig:stacked}
\end{figure*}

While analyzing individual absorbing clouds can tell us about the processes taking place within these few examples, stacking the spectra of many IGM clouds allows us to better quantify the average strength of the lines.
We mean stacked the 341 IGM clouds, both detections and nondetections, in \Ds with a redshift that allows \CIVs and \OVIs to be observed, with the results being shown in Figure~\ref{fig:stacked}.
The spectra was centered on the $z$ of the \Lyas before stacking.
After stacking all IGM clouds, we normalized the continuum using polynomials of 2nd or 3rd order. 
The errors from individual spectra were added in quadrature to obtain the stacked spectra errors, which we show in the bottom panel of Figure~\ref{fig:stacked}.

The log$N$ and $v_\mathrm{obs}$ of the stacked spectra were constrained using the Adaptive Optical Depth (AOD) method \citep{Savage1991,Lehner2020}, which we were able to use since the features included in the stacks were in the linear region of the curve of growth.
The IGM line lists published by \citet{Danforth2016} contain a flag indicating whether or not the line is saturated, which we use to confirm that the clouds included in the stacks are in the linear region of the curve of growth.
The AOD method uses the normalized flux in velocity space to estimate the apparent optical depth in each pixel such that $\uptau_a(v) = \mathrm{ln}\left[F_\mathrm{cont}(v)/F_\mathrm{obs}(v)\right]$.
Since we are in the linear region of the curve of growth, the apparent column density in each pixel can be found by assuming the absorber is unsaturated such that $N_a(v) = 3.768 \times 10^{14}\ \uptau_a(v)/(f\lambda [\mathrm{\AA}])$ $\left[\mathrm{cm}^{-2}\ (\mathrm{km\ s^{-1}})^{-1}\right]$.
This gives $N$ by integrating over the velocity range of the absorber.
We define the width of the stacks as half the velocity width where the normalized flux is 1$\sigma$ below the continuum ($V_{1\sigma}$) after binning the spectra by 4 pixels.
These results are presented in Table~\ref{tab:stack}.

Weak absorption was detected in the \CIVs and \OVIs spectra at $>$5.5$\sigma$.
No features were measured in \SiIVs or \NV, even with the higher signal-to-noise ratios.
As as result, we report the upper limits for these ions as twice their associated uncertainties within $\pm$50~\kms.
While it appears by eye that narrow features are present in \SiIVs and \NV, they were not detected at 2$\sigma$.
It may not be surprising that we do not detect \SiIVs or \NVs in the stack given that the strength of an absorption line is dictated by the product of the elemental abundance, the fraction of the element in the ionization state, and oscillator strength of the line.
For the strongest transitions of \CIVs and \OVI, the product of their abundances and oscillator strengths are similar while the product for \SiIVs and \NVs are an order of magnitude smaller (see Table~4 from \citet{Morton1988}).
Thus, \SiIV~/~Si and \NV~/~N would need to be $\gtrsim$10 times \CIV~/~C and \OVI~/~O for each ion to produce profiles of similar strengths.

The AOD method reveals that, on average, \CIVs and \OVIs are kinematically aligned with their associated \Lyas absorbers, with $\left|v_\mathrm{obs}\right| < 20$~\kmss in both stacks.
From the individual absorping clouds we have analyzed, we believe this trend would continue in \SiIVs and \NVs if the absorption was stronger.

We constrain the distribution of $v_\mathrm{obs}$ of the clouds that went into the stacks by measuring $V_{1\sigma}$ for the combined features.
The widths of \CIVs and \OVIs suggest hotter ions are more likely to differ kinematically from their \Lyas absorbers. 
We hypothesize that these are caused by gas kinematics responsible for non-equilibrium processes producing the coronal transitions.

As seen in Figure~\ref{fig:ionize_all}, the column density ratios of the stacked spectra are consistent with only SI models.
In addition, these ratios are inconsistent with PIE and CIE models, suggesting that the ionization of the IGM is frequently driven by non-equilibrium processes.
This shows that while SIs do not play a major role in the individually absorbing clouds we have analyzed, they may play a part in the overall ionization of the IGM. 
However, it is worth noting that the stacked spectra are not suitable for ratio studies and these results may suffer from issues with averaging multiple populations as one.

\begin{deluxetable}{ccccc}
\centering
\tablecolumns{5}
\tablewidth{0pt}
\setlength{\tabcolsep}{2pt}
\label{tab:stack}
\tablecaption{Summary of Stacked Spectra}
\tablehead{
\colhead{Ion} & \colhead{$W \left[\mathrm{m\AA}\right]$} & \colhead{log$\left(N/\mathrm{cm}^{-2}\right)$} & \colhead{$v_\mathrm{obs} \left[\mathrm{km\ s}^{-1}\right]$} & \colhead{$V_{1\sigma} \left[\mathrm{km\ s}^{-1}\right]$} \\
\colhead{(1)} & \colhead{(2)} & \colhead{(3)} & \colhead{(4)} & \colhead{(5)}}
\startdata
\SiIV & $<$1.92 & $<$11.34 & --- & --- \\
\CIV & $13.01 \pm 2.47$ & $12.51 \pm 0.05$ & $-0.8 \pm 30.58$ & $43.0 \pm 21.5$\\
\NV & $<$1.15 & $<$11.73 & --- & ---\\
\OVI & $16.88 \pm 1.43$ & $13.13 \pm 0.04$ & $-18.5 \pm 12.91$ & $85.9 \pm 21.5$\\
\enddata
\tablenotetext{}{Column~(1) shows the ion being stacked. Column~(2) is the equivalent width of the stacked spectra. Columns~(3) and (4) are the column density and velocity centroid of the stacked spectra measured through the AOD method in units of cm$^{-2}$ and \kms, respectively. Column~(5) indicates the 1$\sigma$ width of the stacked spectra in units of \kms.}
\end{deluxetable}

\section{Summary} \label{sect:summary}

In this work, we explore the ionization processes responsible for the ionization of the WHIM phase of the IGM in the local universe by analyzing absorbing clouds with \CIV, \NV, and \OVIs detected at $\geq$3$\sigma$ from \D.
Our results are summarized as follows:

\begin{enumerate}
    \item Only one cloud was found to be in equilibrium (system 5), specifically CIE with a temperature near 10$^{5.3}$~K. This sight line is also only $\sim$16~kpc from a star-forming dwarf galaxy, hinting that the higher densities found closer to galaxies allow the diffuse halo gas to cool faster than it does further away, where the densities are lower.

    \item None of the clouds were found to be consistent with PIE models, even though the IGM is frequently assumed to be in PIE. Each absorbing cloud required multiple densities to reproduce the observed abundances. In particular, these models were not able to reproduce the amount of \CIVs and \OVIs we see without needing significantly more \NVs than was found.

    \item We compare our observations to four non-equilibrium models: SI, RC, CI, and TML. No system was found to be consistent with the expected ratios from SI or RC models. The system towards PG~1216+069 at $z = 0.12389$ (system 3) is consistent with the expected values for CIs when comparing to \CIVs and \SiIV; however, the system towards PG~1116+215 (system 1) and the one at $z = 0.12360$ towards PG~1216+069 (system 2) only match when comparing to \CIVs and \SiIV, respectively. System 2 is also on the boarder of the predicted ratios of the TML models in the left panel of Figure~\ref{fig:ionize_all}, suggesting they are likely contributing to its ionization state in addition of CIs.

    \item The ionization models we explore here cannot reproduce the ratios of the cloud along the PG~1424+240 sight line (system 4). We note that this system is near the boarder of CIs and TMLs in both panels of Figure~\ref{fig:ionize_all}, which could indicate that these mechanisms are playing some role in the observed ionization state. However, other processes may also be in play since the feature is narrow well beyond the COS line spread function. The true uncertainty in column density is large.

    \item Stacking the spectra of all absorbing clouds within the redshifts that allows for the simultaneous detection of \CIVs and \OVIs revealed faint absorption features at $>$5.5$\sigma$ in \CIVs and \OVI. However, \SiIVs and \NVs were not detected at 2$\sigma$. The column density ratios of these stacks are consistent with SI models, suggesting that SIs may be another prominent ionization mechanisms in the IGM, even if not in the individual clouds presented here.
\end{enumerate}

From these it is clear that the IGM in the local universe is ionized primarily by non-equilibrium processes, in particular in the outer CGM and IGM.
Further investigation of these three coronal lines (\CIV, \NV, and \OVI) in star forming galaxies, with HST programs such as COS-MAGIC (HST-GO-17093), will shed more light on the processes which drive the ionization of the CGM and IGM.

\begin{acknowledgments}
We thank the anonymous referee for the helpful comments which improved the paper.
B.K. and S.B. were supported by NASA/STScI through grants HST-GO-17093 and NSF grant 2108159.
B.K. would like to thank Tyler McCabe for their fruitful discussions regarding this project.
B.K., A.R., and S.B. acknowledge the native people and the land that Arizona State University’s campuses are located in the Salt River Valley.
The ancestral territories of Indigenous peoples, including the Akimel O'odham (Pima) and Pee Posh (Maricopa) Indian Communities, whose care and keeping of these lands allows us to be here today.
\end{acknowledgments}

\facilities{HST (COS); MAST.}

\software{\texttt{matplotlib} (v3.2.2;~\citealt{Hunter2007}), \texttt{astropy} (v4.2.1;~\citealt{Astropy2013,Astropy2018}), \texttt{CLOUDY} (v.17;~\citealt{Ferland2017}), \texttt{numpy} (v1.22.0;~\citealt{Harris2020}), \texttt{scipy} (v1.6.2;~\citealt{Virtanen2020}), \texttt{pandas} (v1.3.5;~\citealt{Reback2021}).}

\bibliographystyle{aasjournal}
\bibliography{References}

\appendix

\begin{figure*}
    \centering
    \figurenum{A1}
    \includegraphics[width=\linewidth]{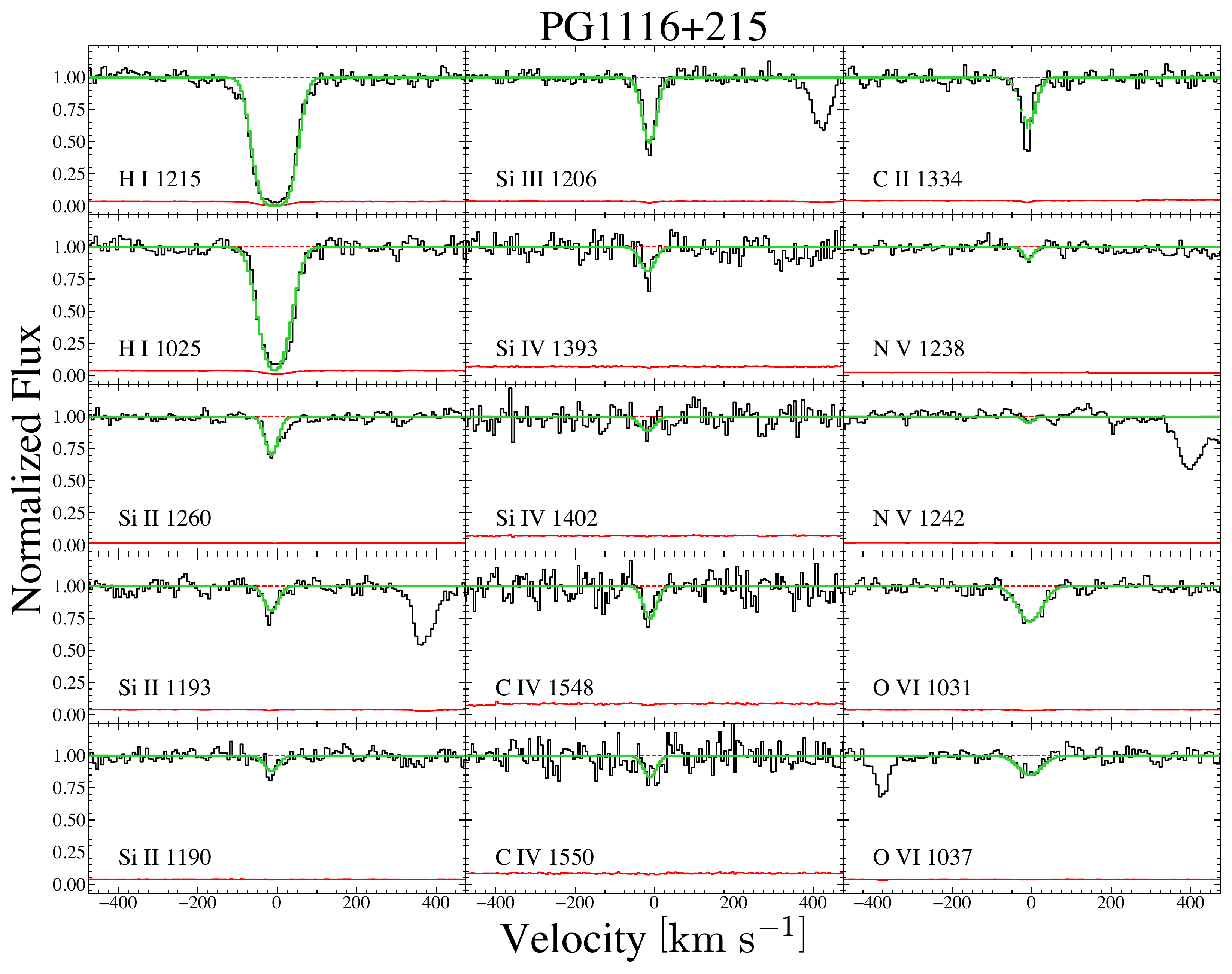}
    \caption{Normalized spectrum of the absorbing cloud along the sightline PG~1116+215. Each panel was centered on the \Lyas velocity of the cloud before fitting. The normalized flux and uncertainties are shown in black and red, respectively. The associated Voigt profile fits are shown in green. Intervening absorbers near $v_\mathrm{sys}$ that are different from the ion being plotted are labeled in blue.}
    \label{fig:spec_PG1116}
\end{figure*}

\begin{figure*}
    \centering
    \figurenum{A2}
    \includegraphics[width=\linewidth]{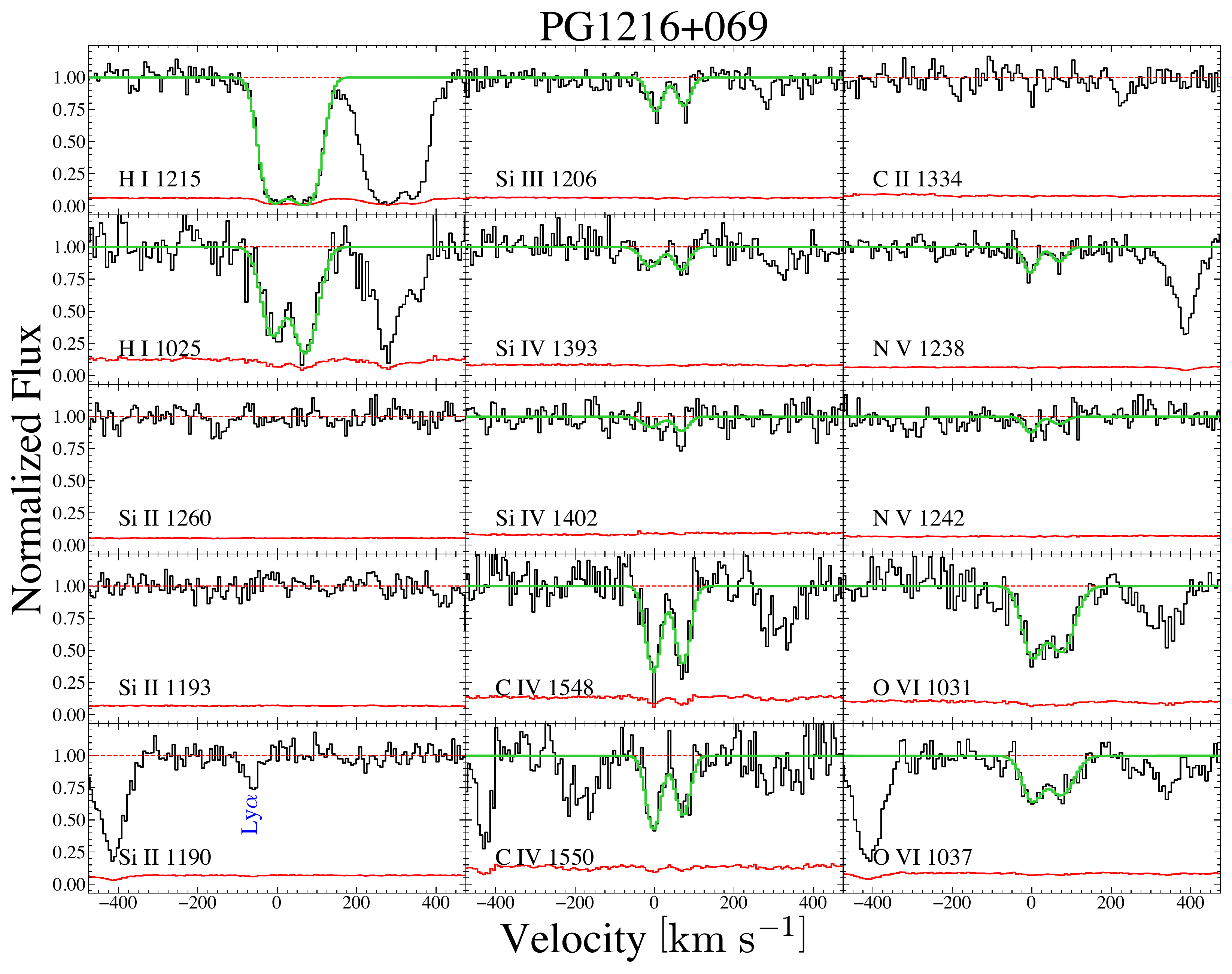}
    \caption{Same as Figure~\ref{fig:spec_PG1116}, but for the absorbing clouds along the sightline PG~1216+069.}
    \label{fig:spec_PG1216}
\end{figure*}

\begin{figure*}
    \centering
    \figurenum{A3}
    \includegraphics[width=\linewidth]{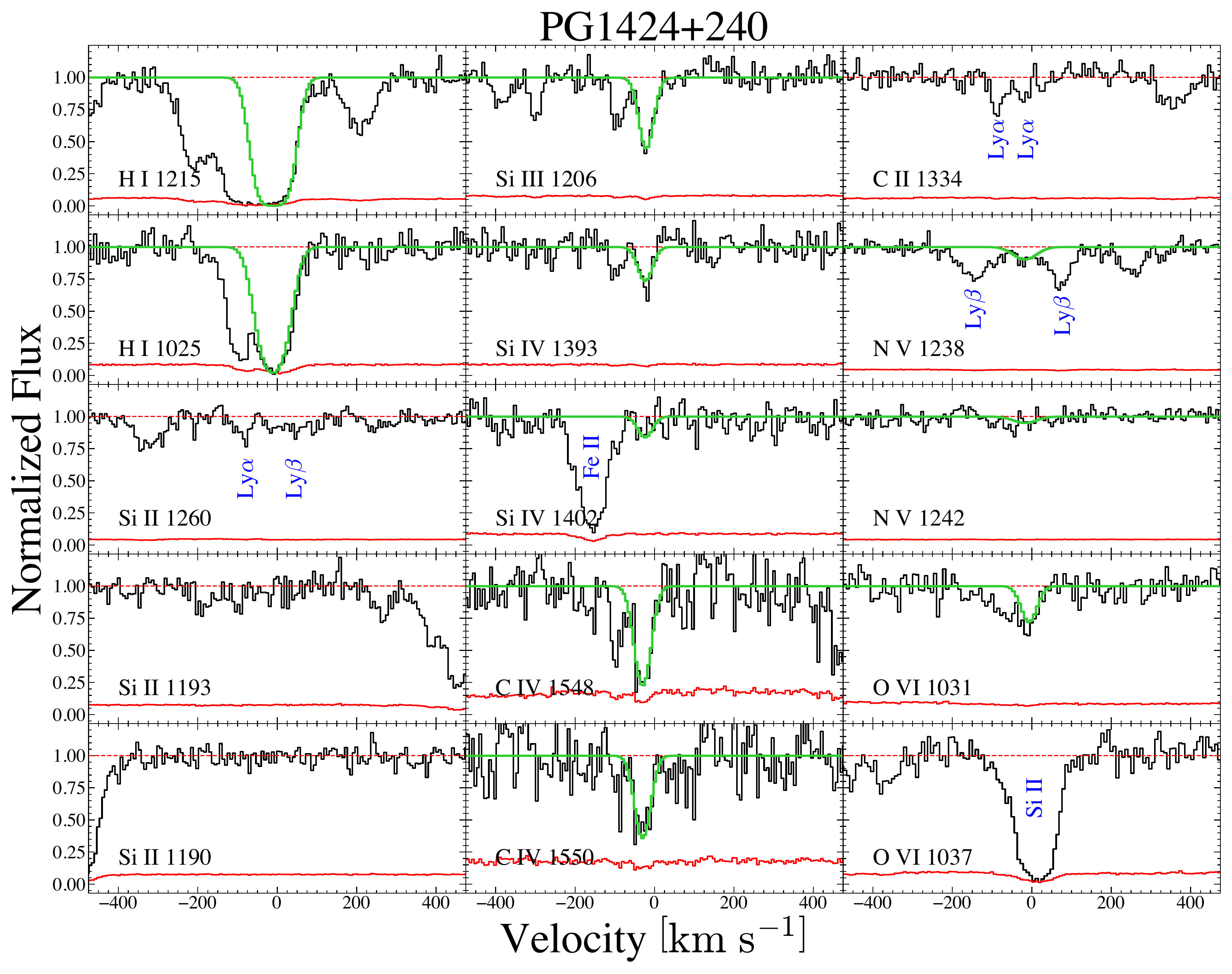}
    \caption{Same as Figure~\ref{fig:spec_PG1116}, but for the absorbing cloud along the sightline PG~1424+240.}
    \label{fig:spec_PG1424}
\end{figure*}

\begin{figure*}
    \centering
    \figurenum{A4}
    \includegraphics[width=\linewidth]{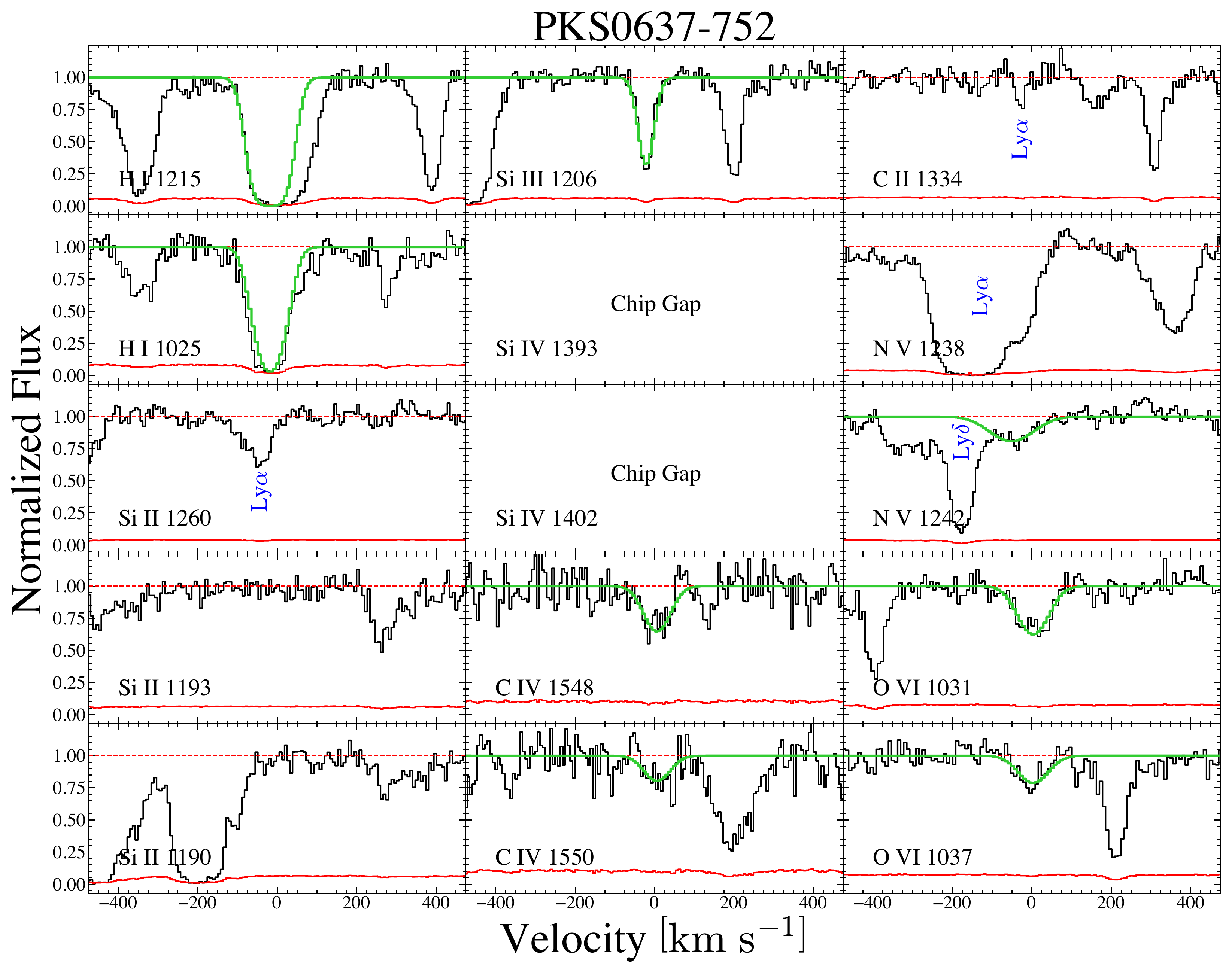}
    \caption{Same as Figure~\ref{fig:spec_PG1116}, but for the absorbing cloud along the sightline PKS~0637-752.}
    \label{fig:spec_PKS0637}
\end{figure*}

\end{document}